\documentclass[12pt,nofootinbib]{revtex4-1}
\newcommand{\ave}[1]{\left\langle #1 \right\rangle}

\newcommand{\eqcomma}{\phantom{AA},\phantom{AA}}

\usepackage{graphicx}

\begin{document}
\title{Quantum magic: A skeptical perspective}
\author{Giorgio Torrieri}
\affiliation{FIAS,
  J.W. Goethe Universit\"at, Frankfurt A.M., Germany  \\
torrieri@fias.uni-frankfurt.de}
\begin{abstract}
Quantum mechanics (QM) has attracted a considerable amount of mysticism, in public opinion and even among academic researches, due to some of its conceptually puzzling features, such as the modification of reality by the observer and entanglement.

We argue that many popular "quantum paradoxes" stem from a confusion between mathematical formalism and physics;
We demonstrate this by explaining how the paradoxes go away once a different formalism, usually inconvenient to perform calculations, is used.
 we argue that some modern developments, well-studied in the research literature but 
generally overlooked by both popular science and teaching-level literature, make quantum mechanics (that is, "canonical" QM, not extensions of it) less conceptually problematic than it looks at first sight.

When all this is looked at together, most ``puzzles'' of QM are not much different from the well-known paradoxes from probability theory.    Consequently, ``explanations of QM'' involving physical action of consciousness or an infinity of universes are ontologically unnecessary
 \end{abstract} 
\maketitle 
\section{Introduction} All the way from its origins, the theory of quantum mechanics (QM) \cite{qm} has enjoyed a resounding experimental success, but has elicited unease regarding its  philosophical implications, and place as a scientific theory.

A lot of research effort on the part of distinguished scientists \cite{penrose,deutsch,susskind} (founders of QM among them! \cite{heisen,schrodinger,bohr,epr}) has gone into ``interpreting'' quantum mechanics.
This effort has produced quite a few candidates for interpretation, 
ranging from the sensible but ambiguous Copenhagen interpretation (``quantum variables only refer to what can be known to us, rather than any objective reality'')
esoteric ideas (such as ``many universes'' and a role of consciousness in quantum physics), less ontologically
troublesome extensions (``hidden variables'')  as well as quite a few ``paradoxes''.

  Some of 
the paradoxes have been very useful at highlighting the trouble that 
ontologically straight-forward extensions of QM would have with 
causality and relativity.  Their net effect, however, has been to convince 
many people that QM has some esoteric dimension which 
places it outside the empiricism, positivism and skepticism on which science is traditionally based.

Thus, the idea that QM assigns a special role to 
consciousness via the ``wavefunction collapse'' is often repeated in the 
popular press. Pseudoscientists, ranging from psychics to astrologers, 
have milked this interpretation for all its worth:  Scientists claim that 
the human mind has some spooky unobservable ``quantum'' influences on 
reality.  So do we.  Ergo, what we do is not so different from what the 
scientists do (see \cite{bleep,jessjournal,koestler}, but we could give many {\em many} more examples).

To be fair, scientists have not been much better.  "Interpretations of QM" relying on 
different universes \cite{everett} or on unfalsifiable "hidden variables" (which, as Bell 
has shown, need to propagate {\em instantaneously} across an arbitrary distance to be consistent with QM \cite{bell}; This means they can 
also travel backwards in time in some reference frame) regularly appear in even research-level journals.
Such ideas should ring all sorts of 
alarm bells among skeptics, yet they are often covered even by science popularization journals as up-to-date scientific research.

Most physicist \cite{coleman}, myself among them, regard these things with skepticism, yet 
very few papers have been written outlining the skeptical case.

Skepticism, in this context, does not mean that QM is not 
real, or not weird and surprising.
It means that QM does not need mysticism to 
be conceptually understood.  To comprehend QM, it is essential to try to disengage ourselves from any ``common sense'' learned from our macroscopic classical world, while at the same time retaining the assumptions basic to skepticism, in particular regarding the world being knowable through empirical observation and testing.   It is not always easy to combine the two, but it is possible, and in the last decades, has been done very well as far as QM is concerned.  
\section{Quantum mechanics: A short introduction}
Quantum mechanics's history, unlike that of the other great revolution in XXth century physics (special and general relativity), is very ``phenomenological'':  There was no leap of understanding of several puzzles in terms of a new universal principle.
Rather, the quantum paradigm imposed itself in a series of steps which were initially somewhat Ad hoc, but were gradually understood to lead to a comprehensive and mathematically elegant theory.

Thus, initially, ``quantum mechanics'' was a set of specific explanations to experimental puzzles.
Planck has shown that, to explain the spectrum of thermal radiation it was necessary to assume that energy was ``quantized'', with the energy of each quantum being proportional to the frequency of the emitted light wave.
Einstein posited that if one considers these quanta as ``particles of light'', one could explain the photoelectric effect, provided that the amplitude of electromagnetic waves is interpreted as the probability density function of the particle to be in a given place in space.
Bohr worked out that to combine Rutherford's picture of the atom with the experimentally observed discrete energy levels, and to render the atom stable against radiation, it was necessary to posit that atomic energies have to be discrete.  

He further realized that this can be elegantly derived by assuming that the ``action'', the integral of the difference between potential and kinetic energy over a classical orbit, be ``quantized'', an integer multiple of a fundamental physical constant.    This constant, as it turns out, is the same as the one used by Planck, suggesting that all these phenomena can be understood in terms of a coherent picture.

Getting to this coherent picture, however, required a lot of work and a new generation of physicists.   The first crucial insight was due to Heisenberg, who realized that quantization followed if one considered ``observables'' (literally, things that experiments could observe) not as simple numbers (as in classical mechanics) but as projections (the technical term is Eigenvalues) of matrices (more abstractly, ``operators'')\footnote{\label{notation}A note on notation: Henceforward, operators distinguished from numbers by using a hat $\hat{A}$ vs $A$.  As we will discuss, operators also correspond to observables probability distributions. Henceforward $\ave{\hat{A}}$ means taking the averaging of this probability distribution  }
The crucial difference is that, while numbers commute ($a b = b a$), with matrices it is not necessarily the case.   A given experimental outcome is equivalent to projecting the physical system in a given state.   Thus, a subsequent measurement will, in principle be different depending whether or not other measurements of this type were previously done.

While this was met with some conceptual puzzlement (the observer seems to irreducibly change reality every time they make an observation), it was possible to rationalize this without the mysticism in vogue today:  As Heisenberg thought, all this meant is that there is an endemic uncertainty in the measurement process, so that, for example, measuring position inevitably meant introducing a disturbance in momentum, while measuring a time meant disturbing the energy.  Hence, the error bars of the two combined measurements had to be larger than a constant, that happened to be equal to Planck's constant\footnote{Here $\Delta A=A-\ave{A}$.  See footnote \ref{notation} for notation }
\begin{equation}
 \ave{\Delta \hat{p}}\ave{\Delta \hat{x}} \geq \frac{\hbar}{2}   \eqcomma \ave{\Delta E}\ave{\Delta t} \geq \frac{\hbar}{2} 
\label{uncert}
\end{equation}
Eq. \ref{uncert} is of course a problem if you want to know everything about everything, but there is no need to invoke consciousness and action at a distance as yet.
Note that this principle is a {\em consequence} of the assumption that physical observables such as position and momentum are represented by matrices that do not commute \cite{qm}
\begin{equation}
\label{uncertprinc}
\hat{x}\hat{p} - \hat{p}\hat{x} = i \hbar
\end{equation}
The appearance of an imaginary number is intriguing, but one has to remember that any observable quantity will always be real, as long as the matrices representing physical observables obey a mathematical property called ``hermiticity'' (see section \ref{formalism}).

Heisenberg's picture was still somewhat counter-intuitive and, more importantly for physicists, devilishly difficult to compute with.
Enter Schrodinger, who showed that the exactly the same physics can be derived from assuming, like Einstein did, that probabilities of finding particles are described by the dynamics of a wave-like equation.   The quantized energy levels,and the equivalent of Heisenberg's ``states'',  correspond to the harmonic modes of a wave.

\begin{figure}
\includegraphics[scale=0.2]{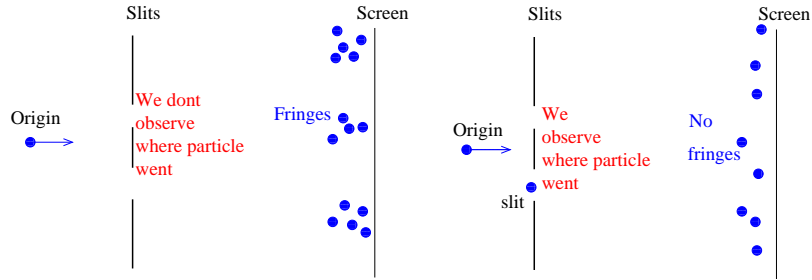}
\caption{\label{slits} The results of the double slit experiment depending on whether the observer watches where the particle goes between the two slits or not}
\end{figure}

The Schrodinger's picture is both mathematically elegant and simple to calculate with.
However, it introduces some additional conceptual issues.  In the Schrodinger's picture the observer does not merely introduce an uncertainty: The {\em act of making an observation} profoundly changes the solution (``wavefunction'') in a way that is in principle completely non-local:  One could invent an experimental setup where making an observation here means changing the wavefunction in the Andromeda galaxy (see next section).
And moreover, the evolution of the wave is radically different depending if the wave ``just evolves'' under the action of a physical system, or if the physical system is ``consciously observing'' the wave.    According to the Schrodinger equation, a particle going through a two-slit experiment behaves like a wave (ie, it goes through {\em both slits at once, and interferes with itself}, thereby generating diffraction fringes) {\em if it is not observed in-flight}.    If its position is observed in-flight to a precision smaller than the slit separation, the particle really behaves approximately as a particle, and the interference fringes disappear (Fig. \ref{slits}.  This has been of course extensively confirmed in experiments).

No wonder scientists, including Einstein and Schrodinger himself, started immediately to invent paradoxes associated with this scenario.    How complicated does a ``quantum particle'' have to be?   Can it just be a microscopic particle whose spin is either up or down, or also a macroscopic, even live object, such a cat (``Schrodinger's cat'') which can be both alive and dead until observed?   And if not, why not?  Where is the macroscopic threshold where common sense starts to prevail?

Since then, QM have been steadily better explored and understood.
Still different ways of deriving the same results have been discovered (among others Feynman's sum-over-paths approach \cite{qm}) and a firm consistent mathematical background in terms of Hilbert spaces has been developed.
QM has also been extended to very complicated, even infinite particle systems, which has given us still new (but remarkably little known to non-specialists) conceptual breakthroughs such as renormalization and decoherence, which I will describe in other parts of this article.
 Armed with all this baggage, QM loses {\em a lot} of its initial ``mystical'' features, while remaining a fascinating theory of nature.
\section{Some quantum paradoxes and their mystical ``explanations''}

Quantum paradoxes, roughly, come in three types, with mixing between the three also generating new paradoxes.

Historically, the first is the Einstein-Podolski-Rosen (EPR) setup \cite{epr}.  Essentially, one takes a correlated system, eg, a particle with zero ``spin'' decaying into two particles of spin, respectively, 1 and -1.
Spin is conserved, and can have three components (let us consider just two, x and y).  However, by quantum uncertainty, measuring the spin in the x direction makes the y direction uncertain.
However, when one measures the spin of one particle of such correlated pair and finds it to be 1, the other pair's particle ``magically'' assumes the certain value of -1, even if that particle is now very far away (the Andromeda galaxy, wherever).
Moreover, an experiment has been performed \cite{aspect} where the {\em decision} which direction to measure has been taken {\em after} the particles were already inflight, yet the correlation kept being there\footnote{This also shows that quantum entanglement is profoundly different from simply Bayesian understanding of correlated probabilities, an argument called ``Bartlemann's socks'' \cite{bell} }.  ``How do the particles know''?  In particular, how do the particles know without information traveling faster than light?  (it should be noted that this phenomenon, as spooky as it looks in a relativistic world, {\em can not} be used to send signals faster than light because the values of the experiment at each end are still random beyond the constraints imposed by the conservation laws).

The particle-wave duality within QM described in the previous section, can be used to generate a whole array of paradoxes and even to do things impossible in a classical world:
  For example, the ``Elitzur-Vaidman bomb-tester'', and more generally interaction-free measurements \cite{elitzur}:  Imagine that we had a bomb that exploded at any contact with its fuse, no matter how light, but we had to distinguish a ``real'' bomb from a ``dud'', given an infinite sample of candidates, just by testing the fuse but without making the bomb go off.   Classically, we would not be able to do it, any contact with the fuse would make the bomb explode.

Quantum mechanically, however, it would be possible: A quantum particle in contact with a dud would behave as a wave, since a dud would not be a measuring device.  With a real bomb, it would behave as a particle.  This can be detected, with interference fringes, {\em even if the bomb does not explode}: Just partially cover one of the slits with the ``candidate'', wait for a few photons to pass, and when no interference fringes are seen on the screen, we have a real bomb (Most likely, the bomb will be hit and explode, but given a ``large'' sample of candidate bombs, we could find a real bomb using this method ).
It sounds ``mystical'', but technological applications of this effect could be out in a few years!

Finally, the very quantum nature of objects is very difficult to reconcile with the classical world of our experience.
If particles can be in a superposition of ``spin 1'' and ``spin -1'', why cant we correlate a particle with spin to a device killing a cat (``Schrodinger's cat'') in a cage, and hence 
making this cat in a ``superposition'' of alive and dead?
Of course, this is a highly incomplete thought experiment, since no physicist would know how to characterize a cat in such a superposition, but still, the question should make us uncomfortable, given living things seem to be either alive or dead.
At least, we need an explanation of why does a ``microscopically quantum'' world looks so classical (cats are either alive or dead!) to us.

These paradoxes have given rise to ``interpretations of quantum mechanics'' which, while popular among some scientists, should make a ``skeptic'' uncomfortable\footnote{by this I {\em do not} mean a skeptic of quantum mechanics, but rather someone with a skeptical worldview of the world in line with the principles outlined, for example, in \cite{shermer2}}.

The most straight-forward such explanation is to take the assumptions of QM at face value and to divide objects in the universe between ``conscious'' (which collapse wavefunctions) and unconscious (which merely evolve them).
Until the day we understand consciousness, such an explanation relies on an unprovable dualism, and hence should be put on the same level as ESP (which a no-less famous figure than Arthur Koestler justified in terms of QM \cite{koestler}) or the existence of the soul.

More ``scientific-looking'' is an explanation in terms of ``hidden variables'': Unobservable physical processes which somehow collapse the wavefunction in non-linear ways when macroscopic objects are near.
A set of such explanations, such as the Ghirardi-Rimini-Weber theory \cite{grw}, Roger Penrose's quantum gravity proposals \cite{gravqm} or Bohms original non-relativistic pilot wave \cite{pilot} are ``scientific'' because they produce experimentally falsifiable consequences (and largely already falsified by experiments such as \cite{aspect}, constraints from astrophysics and condensed matter physics, etc. \cite{pascazio}).
One could always make such hidden variables unobservable and non-causal, but then their existence, for the skeptic, should be apar with the existence of the quantum mind collapsing the wavefunction.

Finally, a class of ideas would want to make QM reconcilable with classical probabilities by postulating the existence of infinitely many universes \cite{deutsch,susskind}, with each outcome encoded in the wavefunction really happening in a different universe.  While this idea is both popular with some physicists and many science fiction writers, until the day we can travel from one quantum universe to the next, each universe, to a skeptic, should have the same validity and heaven and earth.   Stimulating ideas perhaps, but not part of our understanding of the world.

What these ideas have in common is their desire to fit the world to some common-sense assumptions we make about it.   We grew up in a ``classical'' world, we learn that the microscopic world is quantum, and hence weird and surprising, and rather than trying to learn to live with it, we try to make the quantum world ``like the one we grew up with''.

In fact, QM is {\em so} weird and surprising that even the very esoteric interpretations, such as the many universes one, can not capture the 
subtlety normally
associated with quantum paradoxes. For example, theorems such as the Kochen-Specter theorem \cite{kochen} (essentially, a generalization of EPR with {\em three particles}) imply that during its quantum evolution {\em
no} "classical" variables can fully describe the probabilities of a quantum system. The system is {\em not} in a 
yes-no state, even when yes and no are unobserved, in different universes or described by phantom hidden variables.   Microscopic reality really seems to be contextual, the answers depend on the questions you ask\footnote{The Kochen-Specker theorem really nails the inadequacy of classical-looking hidden variable interpretations of quantum mechanics.   Even advocates of hidden variables admit that spin has to be contextual although space and momentum are fixed by a hidden-variable-laden ``quantum potential'' \cite{bookrefgiacosa}.  Of course, the separation between space and spin is only a non-relativistic approximation, and one could invent a K-S type setup with position and momentum.}.

As usual, reality is much more rich with surprises than any speculation trying to fit the world to our prejudices can ever be.
The world is, on a microscopic level, quantum, we should get used to it.
However, we should not abandon our skepticism towards immaterial concepts and 
unfalsifiable ideas when trying to make sense of it.   Can one successfully think in a quantum way {\em and} maintain skepticism? I will try to convince you it is the case.

\section{Is the weirdness in the physics or the maths? \label{formalism}}

A truism familiar to many scientific fields is that all we can learn about the world are statistically inferred probabilities.  Since experiments have always had experimental uncertainties, 
any empirical results are probabilistic in nature (there is always a 
probability  all experiments were wrong). 

It is also a truism that  theories ``put together'' experimental results we have learned in the past:  We {\em do not know} if an apple will go up or down next time it falls from a tree, merely that {\em so far} it has always gone down and we need to seriously revise our theories if it ever goes up.

Given these two constraints, ``all we can quantitatively say about the world'' can be reduced to correlation 
functions (Correlation functions, or correlators, in this contest and in 
the rest of this article, basically mean joint probabilities: The 
correlator between A and B is simply the probability of B given A).   

Furthermore, to a scientist performing experiments ``physical entity''s reality reduces to the ways they can interact with other physical entities.   
Since entities interact by exchanging physical entities (particles, energy, charge, and so on) each interaction invariably changes the entities nature.

No skeptic should be troubled by any of these considerations\footnote{While quantum mechanics is fundamentally different from pre-quantum physics, in ways that I hope will be clarified in the next few paragraphs, these considerations and the resulting uncertainity principle {\em can} be understood by classical arguments: Since every measurement requires energy, the detector has to disperse some energy as heat into the system being studied, with a lower limit of this heat given by Carnot's principle.   Equivalently, as each measurement lowers entropy by obtaining information about the system, some entropy must be created to compensate this, and in information theory this entropy can be expressed as an uncertainity relation between two observables.  Relations like Eq. \ref{uncert} can be derived rigorously with such arguments. Eq. \ref{uncertprinc} and Eq. \ref{heis}, however, although they can also be used to derive Eq. \ref{uncert}, contain more physics than this.  }.
Yet, physicists seldom thought about this until QM, and hence have formulated classical physical laws in deterministic form unless the required theory is explicitly probabilistic in nature (``stochastic'').  

QM can not be described this way, since the correlation functions of different physical results are not independent (the wider the correlation function in position space, the narrower the one in momentum space).  Hence, when we do calculations, we need to consider each measurement's result as a probability distribution, with  previous observations into account in its definition.

What this means is that not just the result, but the ``variance'' and higher cumulants (skewness, Kurtosis, and so on: An infinite number of moments and cumulants) of the probability distribution of the result obey their own equation of motion.     They do not simply come from an uncertainty in initial conditions: They are dynamical degrees of freedom which {\em evolve} just like the expected value of the observable.

In the Heisenberg picture this equation simply states that the rate of change of a given ``operator'' $\hat{A}$ (the matrices discussed around Eq. \ref{uncertprinc}) with time is equal to the commutator with the Energy operator (known as the Hamiltonian $\hat{H}$) multiplied by the imaginary unit $i$ \cite{qm} (see footnote \ref{notation} for notation)
\begin{equation}
\label{heis}
\frac{d}{dt}\ave{ \hat{A}^n} = \frac{1}{i\hbar} \ave{  \hat{A}^n \hat{H} - \hat{H} \hat{A}^n }
\end{equation}
$\hat{A}$ can be the average position and momentum ($n=1$), but also its squared expectation ($n=2$, and the statistical variance is  $\ave{\hat{A}^2} - \ave{\hat{A}}^2$ ), its cubed ($n=3$, and the skewness can be similarly expressed in terms of cubes etc).  Because, due to the relation in Eq. \ref{uncertprinc} typically $\hat{A}$ and $\hat{H}$ do not commute ( The Hamiltonian generally depends on both position $\hat{x}$, through the potential energy and momentum $\hat{p}$, through the kinetic energy), these equations of motion mean that $\ave{\hat{A}^n}$ obeys very different equations of motion from $\ave{\hat{A}}$.   Hence, while for the average of typical experimental observables Eq. \ref{heis} reduces to the classical equations of motion (this is known as the ``Ehrenfest theorem'' \cite{qm}),  the variance and higher cumulants of that observable will never go away and obey their own, highly non-trivial dynamics.

Nothing said so far should be fodder for paradoxes and skeptical puzzlement: The world is uncertain, measurements introduce errors, and eq. \ref{heis} is a way of taking this into account in our calculation.   The bad news is that, since we need {\em all} cumulants to fully take the probability distribution into account, calculating everything this way means solving from initial conditions an infinite array of equations.  This is, at best, very hard.

 For computational ease, we invented functional tools like the ``wavefunction'', where {\em all moments and cumulants} are included.   Essentially, we construct a set of functions $\Psi$ (and their complex conjugates $\Psi^*$) such that, in equation \ref{heis}
\begin{equation}
   \ave{\hat{A}^n} \Leftrightarrow \int  \Psi^*(y) \hat{A}^n(y) \Psi(y)  dy \eqcomma \ave{\hat{H}} \Leftrightarrow \int  \Psi^*(y) \hat{H}(y)  \Psi(y) dy 
\label{shrod}
\end{equation}
where the integration variable $y$ can be either position or momentum.
after some mathematics \cite{qm} \footnote{it can be shown that in this representation momentum $\hat{p} \rightarrow i \hbar \frac{d}{dx}$ if y is position, and position $\hat{x} \rightarrow -i \hbar^{-1} \frac{d}{dp}$ if $y$ is momentum}, Eq. \ref{heis} reduces to a differential equation in terms of $\Psi$.   Once all solutions  $\Psi_i$ are found, all moments and cumulants can be found by integrating (Eq. \ref{shrod}).  This is, essentially, the Schrodinger approach to quantum mechanics.

The ``classical'' analogy to $\Psi$ would be the probability distribution function.  However, some subtlety is present here:  The alert reader will notice an imaginary number $i$ in Eqn \ref{heis}:   In the Heisenberg picture {\em any} physical correlator, evolved from real initial conditions, will end up being real at all time, since, due to the $i$ in Eq. \ref{uncertprinc}, any non-zero term in Eq. \ref{heis} will have two imaginary numbers and hence will be real.
Thus, the imaginary number's role is to take care of a {\em phase} in the systems evolution.

However, translated into the language of Schrodinger's probability generating functions, it will mean that to get a real answer the wave amplitude $\Psi$ needs to be {\em squared} ($\Psi$ needs to be a complex function due to the $i$ but of course all observables are real.  Hence, eqs \ref{shrod} are all in terms of $\Psi \Psi^*$), even through, when these relations are plugged into Eq \ref{heis}, one gets equations of motion for $\Psi$ and $\Psi^*$ separately.   This is quite unlike anything in classical probability.  {\em However}, it is a formalism dependent feature \footnote{It can be shown \cite{rovelli}, in fact, that the squaring rule is a consequence {\em only} of the assumption that, given any physical system, we can ask a continuus infinity of experimental questions but can only obtain a finite amount of data from it experimentally.  These assumptions are consistent with those used to motivate the Heisenberg picture at the beginning of this section}, it only arises in the Schrodinger picture: Eq. \ref{heis} needs no squaring.

The paradoxes of the first two types described in the previous section arise, essentially, from this requirement of squaring the wavefunction to produce a probability (and hence the presence of quantum interference) together with the collapse of the wavefunction at observation: When $\hat{A}$ is observed, $\Psi$ ``instantly changes'' to the solution of the differential equation $\hat{A} \Psi = \lambda \Psi$, where $\lambda$ is a numerical constant corresponding to the result of the observation.   

To understand how puzzling the collapse of the wavefunction is, one needs to remember that paradoxes of this type arise {\em also} in classical probability.
Let us take a famous probability-based puzzle \cite{vos}: The ``Monty Hall problem'', introduced in the appendix and in \cite{vos}.

In the Monty Hall problem, the probability suddenly switches from $1/3$ to $2/3$ when the player (``observer''?) chooses to switch boxes.  Any function characterizing these probabilities would correspondingly change.
For example, assuming $x=1$ means a win and $x=0$ a loss,  the expectation value would change from $1/3$ to $2/3$, while the joint distribution function, 
$G(z) = \sum_{possibilities} p(x) z^x$ changes from  $z/3  + 2/3$ to $2z/3 + 1/3$ (if you choose not to switch,the generating function remains unchanged).   

No one, however, would interpret this as some sort of spooky action at a distance: The generating function is a mathematical object,constructed to calculate probability amplitudes better (rather useless in this problem, but useful for more complicated ones).   It has no physical reality, it is an {\em instrument}.

   In QM, $\Psi$ is not a classical probability generating function, since it has to accomodate the commutation rules of QM.   But, in common with the classical probability generating function, it is not a physical object but rather an instrument.   The fact that {\em all} calculations can be translated into equations of motion for probabilities via Eq. \ref{heis}, where no squaring and interference exists, is the best demonstration for this.

``sudden switches'' of probability amplitudes and correlations when the system is observed are actually quite common in any theory involving correlators and probabilistic systems: Consider the simple Drunkard's walk problem \cite{drunk}, described in the appendix. 
It is immediately obvious that we can not know the drunk's motion with certainty, because, at its core, his motion is probabilistic.  All we can hope to know is that {\em given} the drunk is at  street lamp $i$ at a given time $t_0$, what is the probability the drunk will be at street lamp $j$ at a given time $t_1$ (and higher order functions: what is the probability he will visit lamp $j,k,...$ at times $t_1,t_2,...$).

This is what one means by a correlator, and its the same kind of quantity the mathematical apparatus of QM is meant to calculate.  Once we ``observe'' the drunk to be at a certain street lamp, we ``localize'' his probability distribution to a sharp peak, and all correlators to one $1$ (the drunk is certain to be at a certain streetlamp) or $0$ (and nowhere else).   In time, however, the uncertainty of his position grows, and the correlators accordingly become less sharp.
This is the same behavior of the cumulants of any observable $\hat{A^n}$ when evolved from Eq. \ref{heis}.   Eq. \ref{shrod}, after the squaring and collapse, is {\em guaranteed to give the same answers} as Eq. \ref{heis} \cite{qm}.

The reader should be careful that we are {\em not} claiming that quantum probabilities are the same as classical ones.   They are not, since the probabilistic nature of quantum physics arises from observer-induced uncertainties rather than the classical limited information.  Phenomena such as the ``wavefunction collapse'', however, have their analogues in classical probabilistic phenomena, because {\em calculational instruments} equivalent to the wavefunction can be defined in classical probability.

Most paradoxes of the interference type (as described in the previous section) disappear when the problem is rewritten in the Heisenberg picture language (the downside is that actually solving the problem becomes generally harder):  When enough cumulants are calculated for the probability distribution of ``where will the photon hit the screen given it was at the origin at time $t_0$?'' (Fig. \ref{slits} left panel), fringes start appearing in the probability density function, not from any interference (nothing was squared when solving Eq. \ref{heis}) but because this is simply how the cumulants of the probability distribution function scale for {\em that initial condition}.   The right panel involves the calculation of a different initial condition (``particle localized at the right slid at $t_1$''), and hence it is not surprising the form of this probability distribution function is different w.r.t. the left panel (no fringes form).

Likewise, while it is true that QM would make the Elitzur-Vaidman bomb-test possible \cite{elitzur}, in the language of correlators  the cause of this is simply that the correlation ``the bomb explodes given we have a photon somewhere in the system'' is different between a real and a dud bomb.
This is certainly true, and has non-trivial consequences given quantum uncertainties, but again the paradox becomes less than paradoxical when rewritten in Heisenberg formalism \footnote{Furthermore, if this calculation is done consistently, including quantum field theory corrections, the bomb will explode ``spontaneously'' even if no photon interacts with it due to vacuum fluctuations}.

When correlators are calculated in EPR \cite{epr,bell,aspect} type setups via Eq. \ref{heis}, some  sets of outcomes will always be constant due to conservation laws (spins, independently, from the measured direction, have to be opposite).  Nothing ``collapses'' at the point of measurement over large distances.

Thus, one should be 
careful to not amplify the problems of QM 
by confusing ``formalism'' and ``physics''.
Squaring of the wavefunction $\Psi$ is {\em required by our desire to plug all cumulants into a generating function}.  The generating function must contain the correlations between cumulants contained in the uncertainty relations.  Squaring of a complex number accomplishes all this.
This, however, is a feature of the mathematics of uncertainty relations, not of physics, for one can calculate everything one needs without defining $\Psi$ using Eq. \ref{heis}.


The distinction between quantum and classical also appears when identical particles, or identical experimental outcomes, have to be considered.  \cite{deutsch} argues that ``quantum particle fungibility'' is the key criterion to distinguish quantum from classical mechanics, and illustrates the need for a multiverse understanding of quantum problems.
While many-particle exchange in the quantum world does have some qualitative features unseen in classical physics (the overall $-1$ phase in exchanging fermions \cite{qm}), ambiguity in probability calculations with fungible outcomes is also common in classical probability.  Consider the ``boy or girl'' paradox, similarly to the Monty Hall problem discussed in the popular press\cite{vos2} and introduced in the appendix.
The probability of ``both children being boys if one is a boy'' is 2/3 if the 
children are completely indistinguishable except by gender or 1/2 if the children are somehow 
distinguishable. For normal life, such a distinction is silly, which is 
one of the reasons this problem was so confusing, but in quantum 
mechanics indistinguishable particles are a real possibility, and the 
``two answers'' find a physical application in, say, the branching 
ratios of particles (a $\rho$ particle decays in two charged pions with 
probability 2/3, and into neutral pions with probability 1/3.  If the 
pions were classically distinguishable, all probabilities would have 
been 1/2 ). 
The fact that part of this fungibility arises in "normal" probability means that we need to be careful whether what we consider magic comes from the {\em quantum} or merely the {\em probabilistic}.

 It should not be too surprising that probability gives rise to such 
puzzles.  After all, defining probability and randomness in a rigorous 
way is still largely an open problem \cite{knuth}.  The best definition 
we have involve limits defined on infinite sets of identically prepared systems on which we perform measurements with no prior knowledge about each particular system \cite{knuth}. Note the 
similarity to the ``many-universes'' interpretation of QM \cite{everett}.
The difference,  of course, that no one would suggest that the ``infinite number 
of coin tosses'' required to define the $50\%$ probability of a coin toss producing ``head'' are somehow really happening in different universes.
Perhaps, arguing about many universes in quantum mechanics is equally pointless.
\section{So why all the controversy?}
The considerations made in this section should {\em not} be controversial, since the equivalence of the Schrodinger and the Heisenberg picture are studied in most undergraduate quantum mechanics courses.    The Schrodinger picture, and ``quantum states'', are however so much more widely used that many scientists have gotten used to think of doing quantum mechanics as equivalent to determining and manipulating quantum states, and hence of considering states as ``physical predictions'' of QM \cite{faketheorem}.   

It is true that any quantum system can be viewed as a superposition of Hilbert space states decribed by wavefunctions in the Schrodinger picture.   This is, however, not always the most ``useful'' description. For a system whose classical analogue undergoes periodic motion, all observables in the corresponding quantum theory can be easily computed from a set of numbers describing the system's quantum state.  Most quantum systems studied in quantum mechanics courses are like that, but not {\em all} quantum systems.

Consider the quantum analogue of a classically chaotic system.  If one evolves its quantum state according to Schrodinger's equation, one will always get a linear equation of motion.   Evolving Eq. \ref{heis} for {\em any} observable will in general yield a highly non-linear, chaotic evolution, in line with the fact that the system is classically chaotic.

Is this paradoxical? According to our understanding (the widely believed ``Berry's conjecture'') \cite{berry1,berry2}, the Schrodinger's wavefunctions corresponding to such systems can be approximated by pseudo-random functions, highly irregular functions which can not generally be calculated by approximate methods.
By definition, in such a case, knowing the ``state'' of the system is useless, since computing any observable from such a state via Eq. \ref{shrod} will be impossible.    Vice-versa, determining any observable of the system will give an initial condition for that observable's evolution via Eq. \ref{heis}, but will inevitably put the system in a highly non-trivial superposition of such random states.
Hence, Schrodinger's picture and quantum states,while formally valid, become abstractions, useless for extracting information from the system.  Once again, not part of physics but of mathematics.

This insight can be developed much deeper:  As the next section will show, for {\em any} many-particle system (such systems are classically chaotic if the number of particles is more than 3), the correspondence between observables and states inevitably breaks down.   This can be used to solve the remaining ``Schrodinger's cat''-type paradoxes of quantum mechanics, relating to a quantum description of ``macroscopic'' observables.

\section{We do not live in isolation, and life has a measure: A modern view of Schrodinger's cat}

The discussion in the previous section leaves the third type of paradoxes (Schrodinger's cat) open.  To make sense of them, one needs to answer the question of where ``quantum'' ends and ``classical'' begins open.  This is closely related to the question of why does an inherently quantum world look classical at macroscopic scales.

While this is an ongoing active subject of research,we believe that the key ideas and concepts here have been understood, and no modification in QM is required.

First of all what does ``superposition of alive and dead cat'' actually mean?    Superposition of wavefunctions, as we learned in the last section, is a feature of a mathematical formalism used to calculate quantum observables, not of the quantum physics itself.    Assuming we could construct the ``life operator'', it will evolve according to a very complicated version of Eq. \ref{heis}, with the Hamiltonian encompassing all the atoms of the cat with their respective uncertainties.    The question of Schrodinger's cat reduces then to ``could quantum fluctuations transform an alive cat into a dead cat, {\em and vice-versa?}''.   

It happens with all discrete observables, such as quantum mechanical spin (let us assume ``life'' is one: Its either ``on'' or ``off'' just like spin is either ``up'' or ``down'') which are undetermined: typically the equation of motion of the probability of measuring spin up in the x direction given the z direction was measured has an oscillatory solution.  Could something macroscopic like ``life and death of a cat'' oscillate in a similar way?  If not, why not, and where is the boundary? 
Note that entropic arguments (a ``dead cat'' is much less disordered than an alive cat, and hence much easier to make) do not automatically help:  The ``i'' factor in Eqn. \ref{heis} ensures that, provided the microscopic dynamics is time-reversible (as we believe it is when ``living matter'' is considered: The microscopic motion of atoms in organic molecules certainly seems to obey time-reversible laws), the quantum solution {\em must} oscillate between two possible states.   If quantum mechanics can produce a dead cat from an alive one, the reverse must also be possible and have the same transition probability.   While the first possibility is at least conceptually visualizable (the probability better be very small), the second definitely seems outside from our experience.
 
The key issue here is that the very assumptions of QM mean that no continuous observable (energy, position, momentum and so on) can ever be measured precisely.  To measure energy infinitely precisely ( ie, to ensure a particle is in a certain quantum state) would require an infinite time during which the system (``$S$'') can not interact with its environment ($''E''$).
This is of course impossible, except as an approximation.   A very good approximation for atoms but a very lousy one for macroscopic objects continuously interacting with their environment, such as cats.   

For these objects, according to the rules of QM itself, the dynamics does not anymore obey the simple equation \ref{heis}.
Going beyond it, is however very highly non-trivial since, by definition, the environment is a very complicated system, with many continuously interacting particles.  Conceptually, we do not measure the environment, and hence have to average over any quantum observables representing the environment.  Mathematically for an observable $\hat{A}_S$ defined only in $S$ but interacting with the environment $E$, this means modifying Eq. \ref{heis} to \cite{zurek}
\begin{equation}
\label{heisdeco}
\frac{d}{dt}\ave{ \hat{A^n_S} } = \frac{1}{i\hbar} \ave{  \hat{A^n_S}\ave{ \hat{H}}_{E} - \ave{ \hat{H}}_E \hat{A^n_S}}_{S}
\end{equation}
where $\ave{...}_S$ means averaging only the system and $\ave{...}_E$ means averaging only the environment.   Of course, the Hamiltonian of the system generally has both, plus an interaction term $H_{int}$ between them
\[\    \hat{H}=\hat{H}_{S}+\hat{H}_{E}+\hat{H}_{int} \]
Since the system and the environment interact in very complicated ways when $\hat{H}_{int}\ne 0$, quantum observables representing the environment become correlated with quantum observables representing the system (in a way similar to the EPR paradox, which we examined in the last section).   
While this looks like a very particular set-up, all it requires is the existence of an unobserved system $E$ and $\hat{H}_{int}\ne 0$.
The uncertainity principle makes it very likely that these two conditions are satisfied {\em to some extent} for {\em any} system, since 
a given system's position is always smeared out, with a tail extending out into ``the environment'' for any other conceivable system.

Eq. \ref{heisdeco}, unlike Eq. \ref{heis}, is generally both non-linear and time-irreversible, and some quantum observables are stable against such interaction with the environment while some are not.   In ``textbook'' quantum mechanics problems, $\hat{H}_{int}$ is small enough that these instabilities play no role in the evolution of the system.
However, superpositions of systems that look very different macroscopically (eg superpositions between alive and dead cats) are unstable, and hence rapidly decay, leaving classical-looking probabilities (alive {\em or} dead, with {\em no} quantum oscillation) only.  The relative entropy of the system vs the environment makes an appearance here, because initial ordered states are more likely to be disordered {\em once the correlations with the enviroenment} in Eq. \ref{heisdeco} {\em are traced out}.
  
QM remains fundamentally probabilistic, in the sense that until you open the box the cat is ``either alive or dead'' with a given probability.
This is ``true but boring'', since it is indistinguishable from classical probability: One could interpret this in terms of two universes, one with the cat alive and the other dead, but then classical probability is also defined in a similar way, as the previous section and \cite{knuth} show.   Once the measurement is made and the cat is found ``alive'', quantum fluctuations continue existing and being relevant at the subatomic level.   It is however impossible for them to flip the cat into a dead state (or a dead cat into an alive state) because interactions with the environment ensure these fluctuations {\em stay} at the subatomic level and do not influence events at the {\em macroscopic} scale.  

What does ``macroscopic'' mean here?
For a system interacting with a ``hot'' environment, the time between interactions with the environment is of the order of $\hbar/(k T)$, Planck's constant divided by the temperature.   For ``room temperature'', this is of the order of millionth's of a second, much much smaller than the time between a cat's heartbeats.
If one looks at a cat for such a small amount of time once, asking whether the cat is alive or dead is a meaningless question.   Whatever life is, it seems to be an emergent phenomenon arising at timescales of seconds, much longer than the time it takes for the atoms of the cat to interact with the surrounding environment.   Hence, on the scale of lifes processes (say, heartbeats) all  quantum effects seem to have decohered (there is of course a view, advanced by Penrose, that some quantum effects in the brain survive in longer periods and are responsible for producing consciousness.  These have been discussed in \cite{shermerquantum}, and quantitatively in \cite{tegmark}).
For colder systems, this ceases to be true and colder materials do exhibit macroscopic quantum superpositions (superconductivity and superfluidity ultimately derive from this), but, alas, life has not been known to form in such an environment so we can not use it to understand what a superposition between ``alive'' and ``dead'' objects is.    

At room temperature, unless we bombard the cat with very short-wavelength light rays (e.g. gamma-rays, which would again solve the alive-dead controversy by killing the cat),we can only obtain information at scales much larger than $\hbar/(kT)$.  Due to decoherence, these answers will inevitably look classical to an extremely good approximation even if at a microscopic level the cat ``remains quantum'' and its ``wavefunction never collapses''.   

In summary, as long as we are interested in ``long'' processes (we observe the cat over many heartbeats) and ``large'' length-scales (much larger than the inter-atomic separation and the timescale of interactions between cat and medium) and high temperatures (much higher than absolute zero), decoherence ensures that the cat will always look classical even if its microscopic constituents are quantum.  Altering these conditions will also make the quantum nature of the cat's constituents emerge, but phenomena such as life seem only to exist when these conditions are satisfied \cite{gendeco}.

The realization that any physical system (atoms,cats,...) has a ``relevant scale'' associated with it was a profound conceptual revolution brought about by QM.  
It actually inextricably follows from first principles of QM when the macroscopic properties of systems with many microscopic constituents are considered:  On one hand, the momentum of each constituent is to some extent undetermined because of the uncertainity principle, and the momentum uncertainity is the greater the more localized is each microscopic particle.   On the other, {\em total} momentum is conserved, so the momenta of each particle are correlated (entangled): Measuring one will generally determine that of other particles.      This means that when macroscopic properties of the system are known, in general the microscopic properties are necessarily highly indetermined.   Determining them would alter the macroscopic nature of the system.  The concept of ``emergence'', or different rules governing systems at different (``microscopic'' and ``macroscopic'') scales, is therefore highly ingrained in the structure of quantum mechanics.   For a closed system, these rules are quantum at all scales, but an open system (in the sense of Eq. \ref{heisdeco}) the macroscopic rules approach the classical ones even when the system is ultimately quantum.

The formal realization of this principle, effective field theory and renormalization, forms a core of our understanding of particle physics (and several Nobel Prizes, such as the one to Wilson, and recently to 't Hooft and Veltman, have been awarded for developing these concepts \cite{nobel}.  Nobel prize winner Steven Weinberg also worked extensively in this field).
Yet, the full significance of this revolution has not, to my knowledge, been effectively communicated to the general public.   Both the general public and many established scientists, in particular, yet have to fully appreciate that it is both a consequence of the quantum impossibility to fully measure the world around us, and a key ingredient to bring QM in line with our common sense classical world.

\section{conclusions}
Nothing written here negates that QM is an extremely subtle 
and fascinating theory.    Its manifestations, some of which could lead to 
profound technological innovations in the future (eg quantum computing and 
cryptography) are fascinating and puzzling.

However, given the persistence of philosophical confusion regarding quantum
mechanics, even among advanced physics students, perhaps the introductory quantum mechanics curriculum could benefit from reorganization: students might start with a classical stochastic models (e.g. the drunkard's walk, which is easy to solve analytically in one dimension) to get them familiar with the concept of correlators as solutions to physics problems.  Afterwards, they should become familiar with Heisenberg's picture, and understand that while the equations of motion of the expectation of an operator are typically classical, these observables will also have probability distribution functions with infinitely many moments each obeying their own equation of motion.  Only then should the Schrodinger picture be studied, emphasizing its calculational advantages.    

N

Interpretations of quantum mechanics (a typical final topic of introductory quantum mechanics courses) can then be studied from the vantage point of this plurality of approaches and motivations.
One interpretation of QM which has always been free of conceptual complications is the statistical one \cite{ballentine}: Quantum 
mechanics, as a probabilistic theory, applies not to ``a system'', but to an ensemble of identically prepared systems.  Uncertainty relations are
then simply limits on the way this ensemble can be cut by experimental constraints, and quantum mechanics is a way of taking these constraints into account within calculations.  Conceptually, this way, QM is no more puzzling than
any other probabilistic theory, and the ``many-universes'' coincide with the many identical systems required to define probability. The biggest limit on this interpretation has been, historically, that''quantum interference'' seems difficult to reduce to an ensemble of probabilities.  As this article clarifies, however, these are all features of the {\em calculational formalism} rather than the {\em physics}.  Similarly, decoherence  can provide a way of solving conceptual ``Schrodinger's cat type'' paradoxes in a way which {\em derives} from QM, although doing this for a general system is a problem yet to be fully solved.   These ideas are usually referred to in the literature as the ``consistent histories'' interpretation of QM \cite{consistent}.   

I have found that the main conceptual difficulty to such an approach is surrendering ourselves to an inherently and {\em fundamentally} indeterminate description of the world.   Quantum mechanics, however, is not unique in that.
 In special relativity, the "true" frame of reference is as indefinite as the ``true'' state of the system in quantum mechanics.  What counts is the frame of reference of your meter and clock in relation to the frame of reference of the moving object, just like in Quantum Mechanics what counts are the observables we have previously measured.
That quantum mechanics and relativity share this profound similarity has been noted before \cite{rovelli}.   The difference is that, first of all, the undefiniteness in QM is probabilistic while that of relativity is not.   Secondly, in relativity the indefiniteness is tightly related to symmetry while in Quantum mechanics no such easy interpretation is available (although attempts in that direction do exist \cite{faraggi}).
None of these difficulties, however, should make us reject a priori the idea that some characteristics we regarded as merely {\em indeterminate} are actually {\em indefinite}.   THis would not be the first time it happens in the history of science.

It is ironic that one of Heisenberg's {\em main purposes} in creating 
QM is to create a theory which is founded only on what can 
be observed: As he wrote
"It seems more reasonable to try to establish  a theoretical quantum 
mechanics, analogous to classical mechanics, but in which only relations 
between observable quantities occur" \cite{heisen}.
The bottom line is that QM says just that
 the world can only be understood by ``measuring things about it'', and seeing how these things relate to each other,   and takes some of the consequences of this ``contextuality'' (as this concept is known today) to their logical endpoint.
It certainly was a revolution in our understanding, but it is a revolution that a skeptic should {\em welcome} and {\em appreciate}, for it is based on the tenets of skepticism regarding the material and knowable nature of our world.  
Interpretations adding an unprovable mystical layer to QM, be it related to many universes or consciousness, do a disservice not just to the elegance of QM, but to skepticism.
After all, answering Einstein famous objection with "{\em how do you know} that God would not play dice with the universe?  The evidence seems to say otherwise" is an excellent example in skeptical thinking.
\appendix
\section{Classical probability paradoxes}
\subsection{The Monty Hall problem \cite{vos}}
The setup of the problem follows that of a standard game-show:  The player has to choose three boxes, one of which contains a prize, the remaining ones are empty.

However, when the player makes their choice, the game-show host opens one of the boxes the player {\em did not choose},and shows that it is empty (Note that, provided the game-show host knows the right box, and if the number of boxes is three or greater, this is always possible to do).
The player is than offered a chance to switch to the remaining unopened box, or to stick with their choice.

Intuitively, it is not immediately clear why should switching from one unopened box to the other should change any probabilities.  The probability of getting the prize should be $1/3$ for any box.
However, this is incorrect:  The game-show host added more information to the problem when she opened one of the boxes.    When the initial choice was made, the player was not privy to this information: Switching is guaranteed to win you a prize if you chose the {\em wrong} box initially (probability $2/3$), and you have zero probability of getting the prize if you switch from a correct guess (probability $1/3$).   Hence, switching increases the probability from $1/3$ to $2/3$.

This change of probability at switching is very counter-intuitive, which is why this paradox arouses so much confusion when presented in the popular press \cite{vos}.
One way that the reality of this change becomes apparent is to assume that, instead of 3 possible boxes, there are 1,000,000 of them, with only one leading to the prize.  It is obvious that, if after a totally arbitrary choice is made, the game show host opens 999,998 boxes, it is very likely that the {\em only} box left closed is the one with the prize.
\subsection{The boy and girl paradox \cite{vos2}}
Here, the paradox stems from very similar-looking but crucially different questions:
\begin{description}
\item[Question n.1] Mr. Jones has two children. The older child is a girl. What is the probability that both children are girls?
\item[Question n.2] Mr. Jones has two children. At least one of them is a girl. What is the probability that both children are girls?
\end{description}
Intuitively, since the gender of each child is independent of the gender of any other child, the answer to both questions is $1/2$.   The problem is that in question n.2 there are three possible cases: In order of age, the two children could be [{\em girl,girl}],[{\em boy,girl}],[{\em girl,boy}].   In question n. 1 only the first two cases are allowed.
Since each case is equally probable, and in two cases the result is a boy and a girl, the probability of both children being boys is $1/3$ for question n.2 and $1/2$ for question n.1.

Once again, the crux of the ``paradox'' is that probability is determined by the amount of information available to the ``measurer''.   In the first case, the children are ``distinguishable'', and the gender of one of the two is already specified (the case [{\em girl,boy}] is excluded by the available information).   In the second case, this information does not exist and the number of variations changes.   Since each outcome is equally likely, so does the probability.

In the case of children, of course, such a distinction is somewhat spurious (how can children be indistinguishable?), hence this paradox generated a lot of argument even at the level of philosophy.   In Quantum mechanics, however, the distinguishibility of particles is rigorously defineable, and hence deciding which of these two questions is ill-posed is a question that can be objectively answered.
 
\subsection{The drunkard's walk \cite{drunk}}
There is a street, with regularly spaced streetlights, and a drunk who walks from streetlight to streetlight.   
At each streetlight, there is a 50$\%$ chance of him switching directions.
The problem is to analyze the drunk's motion.    Obviously, after $i$ steps, the {\em average} position of the drunk is where he started, since the probabilities of him going in either direction are the same.    The {\em variance}, however, increases linearly with the number of timesteps.

The problem is, of course, entirely classical, but the solutions can only be expressed in terms of conditional probabilities: What is the probability the drunk will be at the $i$th lamp given it was at the $j$th lamp $N$ steps before.
This is the same mathematical objects that ultimately arise out of quantum mechanics calculations.

G.T. acknowledges the financial support received from the Helmholtz International
Center for FAIR within the framework of the LOEWE program
(Landesoffensive zur Entwicklung Wissenschaftlich-\"Okonomischer
Exzellenz) launched by the State of Hesse.
The author wishes to thank Bob McElrath Francesco Giacosa, Ben Koch, Czarina Salido, Jessica Rafka, Michael Lennek, William Nelson, Syksy Rasanen and Christine Nattrass for discussions and corrections


\begin{thebibliography}{10}
\bibitem{bleep} The movie ``What the Bleep Do We Know!?'', based on this premise, was widely praised in the popular press and grossed over 16 million dollars, a very good performance for a low budget documentary.
It was followed by a similar sequel, ``What the BLEEP – Down the Rabbit Hole''
\end{thebibliography}
\end{document}